\documentclass[leqno]{article}
\usepackage{amsmath,amssymb,amsfonts} 
\usepackage{mathtools}
\usepackage{nicefrac}
\usepackage{xspace}
\usepackage{color,graphicx,url,subfigure,fullpage}
\usepackage{tabularx} 
\usepackage{hyperref} 
\usepackage{pdfsync}

\newcommand{\defeq}{\stackrel{\textup{def}}{=}}

\newcommand{\eps}{\epsilon}

\renewcommand{\epsilon}{\varepsilon}

\newtheorem{theorem}{Theorem}[section]

\newtheorem{claim}[theorem]{Claim}

\newenvironment{proof}{\begin{trivlist} \item {\bf Proof:~~}}
   {\qed\end{trivlist}}

\def\FullBox{\hbox{\vrule width 6pt height 6pt depth 0pt}}

\def\qed{\ifmmode\qquad\FullBox\else{\unskip\nobreak\hfil
\penalty50\hskip1em\null\nobreak\hfil\FullBox
\parfillskip=0pt\finalhyphendemerits=0\endgraf}\fi}

\def\qedsketch{\ifmmode\Box\else{\unskip\nobreak\hfil
\penalty50\hskip1em\null\nobreak\hfil$\Box$
\parfillskip=0pt\finalhyphendemerits=0\endgraf}\fi}

\newcommand\R{\mathbb R}

\newcommand{\inparen}[1]{\left(#1\right)}             %\inparen{x+y}  is (x+y)
\newcommand{\cM}{\mathcal{M}}
\newcommand{\st}{\mbox{\rm s.t. }}

\newcommand{\per}{\mathrm{Per}}

\title{{On Convex Programming Relaxations for the Permanent}}
\usepackage{authblk}

\author[1]{ Damian Straszak}
\author[2]{Nisheeth K. Vishnoi}
\affil[1,2]{\small \'{E}cole Polytechnique F\'{e}d\'{e}rale de Lausanne (EPFL), Switzerland}

\begin{document}
\date{}
\maketitle

\begin{abstract}
In recent years, several convex programming relaxations have been proposed to estimate the permanent of a non-negative matrix, notably in the works of  \cite{GS02,gurvits2011unleashing,GS14}.
However, the origins of these relaxations and their relationships to each other have remained somewhat mysterious.
We present a conceptual framework, implicit in the belief propagation literature, to systematically arrive at these convex programming relaxations for estimating the permanent -- {\em as approximations to an exponential-sized  max-entropy convex program for computing the permanent}. 
Further, using standard convex programming techniques such as duality, we establish equivalence of these aforementioned relaxations to those based on {\em capacity-like} quantities studied in \cite{Gurvits06,Gurvits09,AMOV16}. 
\end{abstract}

\section{Introduction}
Let $A\in \R^{n\times n}$ be a square matrix with entries $A_{i,j}$ for $1\leq i,j \leq n$. The permanent of $A$ is defined as:
$$\per(A)\defeq \sum_{\sigma \in S_n} \prod_{i=1}^n A_{i,\sigma(i)},$$
where $S_n$ is the set of all permutations over $n$ symbols, i.e., the set of bijections $\sigma: \{1,2,\ldots, n\} \to \{1,2,\ldots, n\}$.
The matrix permanent makes its appearance in various branches of science and mathematics and algorithms to compute it are sought after, especially in  theoretical computer science.

In a foundational result, Valiant~\cite{Valiant79} proved that it is unlikely that there is an efficient algorithm that computes the permanent of a matrix -- even when the matrix is non-negative (the problem is $\mathbf{\#P}-$complete).   
Jerrum, Sinclair and Vigoda~\cite{JSV04}, building upon a long line of work, showed that the Markov Chain Monte Carlo framework can be deployed to obtain a randomized algorithm to estimate the permanent of any non-negative matrix to within a factor of $1+\eps$ in time that is polynomial in the bit-lengths of $A$ and $1/\eps$.

A different approach to estimate the value of the permanent of a non-negative matrix has been developed -- notably in the works of  Linial, Samorodnitsky and  Wigderson~\cite{LSW98} and Gurvits and Samorodnitsky~\cite{GS02}. 
Roughly, the idea in these works is to reduce the computation of the permanent of a non-negative matrices to that of a $n \times n$ doubly stochastic matrix by computing a certain  {\em scaling}.
The latter work \cite{GS02} reveals how the scalings can be viewed as  solutions to certain  convex programs -- leading to convex programming relaxations for the permanent.
The resolution of the van der Waerden conjecture by Egorychev~\cite{Egorychev81} and Falikman~\cite{Falikman81} -- that the permanent of a doubly stochastic matrix is lower bounded by $e^{-n}$ -- resulted in a deterministic algorithm to estimate the permanent up to a multiplicative ratio of $e^n$. 
Of note here is a relaxation introduced by Gurvits~\cite{Gurvits06}, called the {\em capacity} of the associated polynomial $p_A$ (applies more generally), which allowed him to deploy the theory of hyperbolic or real-stable polynomials to derive the same result.
These ideas were further developed in the works of~\cite{gurvits2011unleashing,GS14} where a curious convex program was presented and, it was shown using an inequality due to Schrijver~\cite{schrijver1998counting}, that this program estimates the value of the permanent up to a $2^n$ factor.

The problem of computing the permanent has also attracted attention in statistical physics and, in particular, the belief propagation literature \cite{WC10,Vontobel13,CY13}.
These algorithms are geared towards computing approximations to the Gibbs distribution associated to the matrix  $A$. 
A bit more formally, if  $G=(V,W,E)$ is the bipartite graph where $|V|=|W|=n$ and $(i,j) \in E$ if $A_{i,j} >0$ and $\cM$ is the set of perfect matchings in $G$, then  the Gibbs distribution induced by $A$ and supported on $\cM$ is one where, for a matching $M \in \cM$, the probability of  $M$, denoted by, $p(M)$, is 
$$ p(M) \defeq \frac{\prod_{(i,j) \in M} A_{i,j}} {{\per(A)}}.$$ 
The idea then is to maintain a proxy to $p(M)$, or a {\em belief} distribution $b:\cM \to \mathbb{R}_{\geq 0}$ and iteratively update it according to certain rules.
It is not clear when and how quickly such algorithms converge although they seem to work well in practice and for locally tree-like graphs.

\medskip
{\em Are these three seemingly distinct approaches to compute the permanent, coming from different areas, connected?}

\medskip
\noindent
The goal of this paper is to present some old and some new results in a unified manner which seem to suggest that there may be a close connection between the latter two approaches. 
The starting point to our work is the well-known fact that the following convex program has $\log \per(A)$ as its optimal value.
\begin{equation}
\begin{aligned}
  \sup ~& \sum_{M\in \cM} b(M) \log \frac{\prod_{(i,j) \in M} A_{i,j}}{b(M)}  \\
\st  
  & \sum_{M\in \cM} b(M) =1\\
& b(M) \geq 0 ~~~~~~~~~~ \forall M\in \cM 
\end{aligned}
\label{eq:exact}	
\end{equation}
Each variable of this program corresponds to a perfect matching in $G$, and the goal is to find a probability distribution over perfect matchings in $G$ that minimizes the Kullback-Leibler divergence from the Gibbs distribution.
Thus, if one could compute the optimal value of this program then one could estimate the permanent exactly.
However,  the number of variables in this program can be exponentially large and it is not clear how to solve it.
One can take the dual of this program, however, that leads (predictably) to the same problem; see \cite{SinghV14} for more details.

We show how we can bypass this barrier and construct approximate convex programs starting from this one by making assumptions on the distribution we are seeking.
Key to these simplifications are some heuristics from belief propagation, which result in  certain relations between these marginals that one (wishfully) expects the optimal distribution $p$ to satisfy.
Assuming these relationships, one can then write the above convex program entirely in  terms of the first order marginals of the distribution $b$ -- dramatically reducing the number of variables.
Of course, as a result of these simplifications, the convex programs obtained are approximations of the one above;  see \eqref{eq:entr} and \eqref{eq:entrg}.
However, as proved by Gurvits~\cite{Gurvits06} and Gurvits and Samorodnitsky~\cite{GS14}, they give us some of the best deterministic algorithms to  estimate the value of $\per(A)$.
We expect that our point of view will lead to new convex programming methods for other counting problems (such as those studied in  \cite{SV16}) by leveraging on heuristic reasonings from belief propagation.

\section{Relaxations via Belief Propagation}
Belief propagation algorithms attempt to find the Gibbs distribution $p$ by maintaining  belief distributions $b$ supported over $\cM$.
These relations are often arrived at by considering the case when $G$ is a tree. 
In the following two subsections, we  show that using two different  relations, we end up with two different convex programs which, in turn, via standard techniques from convex programming allow us to recover the convex programs from \cite{GS02,Gurvits09,gurvits2011unleashing,GS14,AMOV16}.
Some of the results we present are implicit in the papers by \cite{WC10,Vontobel13,CY13,CS15}.
Our contribution is to present the unified picture along with some new relations.

 Let us now introduce some notation to explain the ideas formally. For a probability distribution $b$ supported on $\cM$ we denote by $b_{i,j}(1)$ the probability that the edge $(i,j)$ is present in a random matching drawn from $b$, similarly $b_{i,j}(0)=1-b_{i,j}(1)$ is the probability that the edge is absent. Denote $B_{i,j}=b_{i,j}(0),$ it follows in particular that 
\begin{enumerate}
\item $\sum_{j=1}^n B_{i,j}=1$ for all $1\leq i \leq n$, and 
\item $\sum_{i=1}^n B_{i,j}=1$ for all $1\leq j \leq n$.
\end{enumerate}
In other words, $B=(B_{i,j})_{i,j \in [n]}$ is a doubly stochastic matrix with support determined by $A$.
We denote this set  by $\Omega_n(A)$ and the set of all doubly stochastic matrices by $\Omega_n$. 

\subsection{Gurvits' first relaxation for the permanent}

The starting point here is the  following observation that when $G$ is a tree and $b$ is any distribution over perfect matchings in $G$, then
\begin{equation}\label{eq:prod}
b(M) =\prod_{(i,j): M_{i,j}=1} b_{i,j}(1).
\end{equation}
In other words, the probability of a perfect matching in $G$ is completely determined by the marginal probabilities of the edges and is computed as if all the events (corresponding to picking a particular edge) were independent.
The reason why this is trivially true in the case of a tree is because a tree has at most one perfect matching. Therefore, all the marginals are either zero or one, depending whether a given edge belongs to the unique perfect matching or not. 

The heuristic jump then occurs when one,  wishfully thinking,  hopes that these relations hold for all graphs -- {\em at least when $b$ is the  Gibbs distribution $p$}.
While incorrect, assuming that $b$ satisfies \eqref{eq:prod}, leads us to the convex program as an approximation to  that in~\eqref{eq:exact} in the following manner:

\begin{align*}
\sum_{M\in \cM}  b(M)  \log  \frac{\prod_{(i,j) \in E: M_{i,j}=1} A_{i,j}}{\prod_{(i,j):M_{i,j}=1}b_{i,j}(1)} = &\sum_{M\in \cM}  b(M)  \log  \frac{\prod_{(i,j) \in E: M_{i,j}=1} A_{i,j}}{\prod_{(i,j) \in E: M_{i,j}=1}B_{i,j}}  \\
=&  \sum_{M\in \cM}  b(M)  \left( \sum_{(i,j) \in E: M_{i,j}=1}  \log \frac{ A_{i,j}}{B_{i,j}} \right). 
\end{align*}

\noindent
The expression on the right hand side can be simplified further. Consider

\begin{eqnarray*}
 \sum_{M\in \cM}  b(M)  \left( \sum_{(i,j) \in E: M_{i,j}=1}  \log \frac{ A_{i,j}}{B_{i,j}} \right) & = &    \sum_{(i,j) \in E} \sum_{M\in \cM : M_{i,j}=1}  b(M)   \log \frac{ A_{i,j}} {B_{i,j}}\\ 
 & = & \sum_{(i,j) \in E}  \log \frac{ A_{i,j}}{B_{i,j}}  \sum_{M\in \cM : M_{i,j}=1}  b(M)  \\
 & = & \sum_{(i,j) \in E} B_{i,j}  \log \frac{ A_{i,j}}{B_{i,j}}.
\end{eqnarray*}

\noindent
Thus the relaxation~\eqref{eq:exact} reduces to the problem of maximizing the above expression over $B\in \Omega_n(A)$, i.e., we arrive at the following convex program:\footnote{In~\eqref{eq:entr} we switched from $\Omega_n(A)$ to $\Omega_n$ as the domain for $B$, however it is not hard to see that  the value of the relaxation is not affected.}

\begin{equation}
\begin{aligned}
R_E(A)\defeq \sup_{B\in \Omega_n}     \sum_{(i,j) \in E} B_{i,j} \log \frac{A_{i,j}}{B_{i,j}}.\\
\end{aligned}
\label{eq:entr}	
\end{equation}

\noindent
Using Lagrangian duality, we can now derive the relaxation considered by Gurvits.
For variables $z_1,\ldots,z_n$ consider the polynomial $q_A(z) \defeq \prod_{i=1}^n \sum_{i=1}^n A_{i,j} z_j$.  
One can show using standard techniques that the following is the dual of the above convex program.

\begin{equation}\label{eq:capa}
 R_C(A)\defeq \inf _{z >0} \frac{q_A(z)}{\prod_{i=1}^n z_i}. 
\end{equation}

\noindent
Formally one can prove the following theorem  (see Appendix~\ref{sec:eq} and \cite{CS15}):

\begin{theorem}\label{thm:rc}
For all non-negative $n\times n$ matrices $A$,  $R_C(A)= \exp(R_E(A)).$
\end{theorem}

\noindent
Since the coefficient of $\prod_{i=1}^n z_i$ in $q_A(z)$ is equal to $\per(A)$ it is easy to see that 
$$ \per(A)\leq R_C(A).$$
Furthermore an important result of  Gurvits~\cite{Gurvits06}, which makes this convex program interesting,  asserts that an upper bound holds:
$$R_C(A) \leq e^n \cdot \per(A).$$
Note that even though we arrived at this convex program by making an incorrect assumption on the Gibbs distribution, the result is a relaxation which provably approximates the permanent up to a factor of $e^n$ -- thus yielding a {\em deterministic} algorithm to estimate the value of the permanent of $A$.

\medskip
We conclude this section by showing how \eqref{eq:entr} can be reinterpreted in the language of polynomials, resulting in another relaxation for the permanent implicit in the work of~\cite{AMOV16} (see also~\cite{Gurvits09}).
To explain their relaxation, consider any two positive matrices $C$ and $D$ such that their entry-wise product is equal to $A$, i.e., $A_{i,j}=C_{i,j} \cdot D_{i,j}$ for all $1\leq i,j\leq n$. 
Define $n^2-$variate polynomials $p(x)=\prod_{i=1}^n \sum_{j=1}^n x_{i,j} C_{i,j}$ and $q(y) = \prod_{j=1}^n \sum_{i=1}^n y_{i,j}D_{i,j}$. 
Then the relaxation  is as follows:\footnote{In the relaxation defining $R_P(A)$ one can equivalently optimize over a larger domain for $B$, namely $B\in [0,1]^{n\times n}$.}
$$R_P(A) \defeq \sup_{B\in \Omega_n} \inf_{x>0, y>0} \frac{p(x)\cdot q(y) \cdot \prod_{i,j} B_{i,j}^{B_{i,j}} }{\prod_{i,j} (y_{i,j} x_{i,j})^{B_{i,j}}}.$$
We observe that this relaxation is equivalent to~\eqref{eq:entr}.

\begin{theorem}\label{thm:equiv_ep}
For all non-negative $n\times n$ matrices $A$,  $R_P(A)= \exp({R_E(A)}).$
\end{theorem}

\noindent The proof can be easily deduced from the following observation.
\begin{claim}\label{lemma:poly}
Let $r$ be the following polynomial over $n^2$ variables $x_{i,j}$ (for $1\leq i,j \leq n$)
$$r(x) = \prod_{i=1}^n \sum_{j=1}^n M_{i,j}x_{i,j},$$
where $M$ is any matrix with positive entries. Let $B\in \Omega_n$ be any doubly stochastic matrix, then
$$\inf_{x\in \R^{n\times n}_{>0}} \frac{r(x)}{\prod_{ i,j} x_{i,j}^{B_{i,j}}}=\exp \inparen{ \sum_{i,j} B_{i,j} \log \frac{M_{i,j}}{B_{i,j}}}.$$
\end{claim}
The proof of Claim~\ref{lemma:poly} appears in Appendix~\ref{sec:proof_claim}. 

\subsection{Gurvits' second relaxation for the permanent}
To derive the second relaxation, as before we start with a relation among a distribution $b$ and its marginals which holds (again trivially) when $G$ is a tree.

\begin{equation}
\label{eq:approx2}
b(M) = \frac{\prod_{M_{i,j}=1}b_{i,j}(1)}{\prod_{M_{i,j}=0} b_{i,j}(0)}.
\end{equation}
 \noindent
The reason this is true for trees is that for the unique perfect matching $M$ of $G$ for $(i,j)\in M$ we have $b_{i,j}(1)=1$ and for $(i,j) \notin M$ we have $b_{i,j}(0)=1$.
As in the previous section,  we  make a jump and assume that this relation holds  for any bipartite graph -- at least for the case of the Gibbs distribution. 

Let us  simplify the convex program~\eqref{eq:exact} using~\eqref{eq:approx2} as an assumption.
We obtain that the objective of the convex program~\eqref{eq:exact} can be written as 
\begin{eqnarray*}
 \sum_{M\in \cM}  b(M)  \log \frac{\prod_{(i,j) \in E: M_{i,j}=1} A_{i,j}}{ \frac{\prod_{(i,j) \in E: M_{i,j}=1}B_{i,j}}{\prod_{(i,j) \in E: M_{i,j}=0} (1-B_{i,j})}} & =&  \sum_{M\in \cM}  b(M)  \left( \sum_{(i,j) \in E: M_{i,j}=1}  \log \frac{ A_{i,j}}{B_{i,j}}  + \sum_{(i,j) \in E: M_{i,j}=0} \log  (1-B_{i,j}) \right). 
\end{eqnarray*}
The expression on the right hand side can be simplified as before to obtain

\begin{eqnarray*}
 \sum_{M\in \cM}  b(M)  \left( \sum_{(i,j) \in E: M_{i,j}=1}  \log \frac{ A_{i,j}}{B_{i,j}} \right) & = &    \sum_{(i,j) \in E} B_{i,j}  \log \frac{ A_{i,j}}{B_{i,j}}.
\end{eqnarray*}

\noindent
Similarly, 
$$  \sum_{M\in \cM}  b(M) \left( \sum_{(i,j) \in E: M_{i,j}=0} \log  (1-B_{i,j}) \right) =  \sum_{(i,j) \in E} (1- B_{i,j})  \log (1-B_{i,j}).
 $$
This allows us to rewrite the convex program~\eqref{eq:exact} in the following form: 
\begin{equation}\label{eq:entrg}	
	R_O(A)\defeq \sup_{B\in \Omega_n}  \sum_{(i,j) \in E} B_{i,j}  \log \frac{ A_{i,j}} {B_{i,j}}+ \sum_{(i,j) \in E} (1- B_{i,j})  \log (1-B_{i,j}).
\end{equation}

\noindent
It is worth noting that even though the program~\eqref{eq:exact} is convex, it does not imply that the objective in \eqref{eq:entrg} is convex. 
In fact it is convex only when restricted to the domain $\Omega_n$, which was observed by~\cite{Vontobel13}.
Even more remarkable is the fact that, despite us plugging in the relations between the marginals that are only guaranteed to hold when $G$ is a tree, it was proved that the approximation guarantee is even better as for the first relaxation. 
Here proving that this convex program has any relation to the permanent requires some work and, indeed, Gurvits, using an inequality due to Schrijver \cite{schrijver1998counting},  proved that 
$$\exp(R_O(A)) \leq \per(A).$$
Notice here that the estimate obtained from the convex program provides a {\em lower bound} to the permanent unlike Gurvits' first relaxation.
Subsequently, Gurvits and Samorodnitsky~\cite{GS14}  proved that this quantity in fact provides a sharp estimate for the permanent:
$$  \per(A) \leq  2^n \cdot \exp(R_O(A)).$$

\noindent
Finally, we show how another  polynomial-based relaxation for the permanent that is implicit in \cite{AMOV16} is equivalent to~\eqref{eq:entrg}.  Let, as before, $C$ and $D$ be two positive matrices such that their entry-wise product is equal to $A$, i.e., $A_{i,j}=C_{i,j} \cdot D_{i,j}$ for all $1\leq i,j\leq n$. Define polynomials $p(x)=\prod_{i=1}^n \sum_{j=1}^n x_{i,j} C_{i,j}$ and $q(y) = \prod_{j=1}^n \sum_{i=1}^n y_{i,j}D_{i,j}$. Consider the following
\begin{equation}\label{eq:poly_odd}
R_Q(A) \defeq \sup_{B\in \Omega_n} \inf_{x>0, y>0} \frac{p(x)\cdot q(y) \cdot \prod_{i,j} B_{i,j}^{B_{i,j}} \cdot \prod_{i,j}(1-B_{i,j})^{1-B_{i,j}}}{\prod_{i,j} (y_{i,j} x_{i,j})^{B_{i,j}}}.
\end{equation}
As in the case of the first relaxation, we have the following theorem.
\begin{theorem}\label{thm:equiv_oep}
For all non-negative $n\times n$ matrices $A$,  $R_Q(A)= \exp({R_O(A)}).$
\end{theorem}
This theorem can be easily deduced from Claim~\ref{lemma:poly}. Indeed, for a fixed $B$, $x$ and $y$ are separable in the objective of $R_Q$, hence Claim~\ref{lemma:poly} can be applied separately to both of them, to yield the claimed result.

\bibliographystyle{alpha}
\bibliography{references}

\appendix

\section{Proof of Theorem \ref{thm:rc}}\label{sec:eq}
\begin{proof}
Consider an $n\times n$ non-negative matrix $A$ and assume for simplicity that it is of full support, i.e., all the entries of $A$ are positive. Let us first state the relaxation defining $R_E(A)$ with all constraints explicitly present.
\begin{equation}
\begin{aligned}
	\max_{B\in \R^{n\times n}}  ~~ &   \sum_{1\leq i,j\leq n} B_{i,j} \log \frac{A_{i,j}}{B_{i,j}},&\\
		\st ~~ &  \sum_{i=1}^n B_{i,j}=1,& j=1,2,\ldots, n\\
 &  \sum_{j=1}^n B_{i,j}=1,& i=1,2,\ldots, n\\
	&  B\geq 0.&
\end{aligned}
\label{eq:r_e}	
\end{equation}
Note importantly that the objective of~\eqref{eq:r_e} is concave in $B$. We now derive the dual of the above program. Introduce Lagrangian multipliers $\alpha, \beta \in \R^n$ for the row and column stochasticity constraints. We arrive at the Lagrangian
$$L(B,\alpha, \beta)  = \sum_{1\leq i,j\leq n} B_{i,j} \log \frac{A_{i,j}}{B_{i,j}} - \sum_{i=1}^n \alpha_i \inparen{\sum_{j=1}^n B_{i,j}-1} - \sum_{j=1}^n \beta_j \inparen{\sum_{i=1}^n B_{i,j}-1} .$$
The next step is to compute
$$g(\alpha, \beta) = \sup_{B\geq 0} L(B,\alpha, \beta).$$
To this end let us compute the derivative of $L$ with respect to $B_{i,j}$.
$$\frac{\partial}{\partial B_{i,j}} L(B,\alpha, \beta) = \log A_{i,j} - \log B_{i,j} -1 - \alpha_i - \beta_j.$$
Setting $\frac{\partial}{\partial B_{i,j}} L(B,\alpha, \beta)=0$ gives
$$B_{i,j} = A_{i,j} e^{-1-\alpha_i-\beta_j}.$$
By plugging in these values to the Lagrangian, we obtain
$$g(\alpha, \beta) = \sum_{1\leq i, j \leq n} A_{i,j}e^{-1-\alpha_i-\beta_j} + \sum_{i=1}^n \alpha_i + \sum_{j=1}^n \beta_j.$$
Since the Slater's condition is satisfied for~\eqref{eq:r_e}, strong duality holds and we can deduce that
$$R_E(A) = \min_{\alpha, \beta \in \R^n} \sum_{1\leq i, j \leq n} A_{i,j}\cdot e^{-1-\alpha_i-\beta_j} + \sum_{i=1}^n \alpha_i + \sum_{j=1}^n \beta_j.$$
Finally, we can eliminate one set of variables due to a simple form of the dual objective $g(\alpha, \beta)$. Indeed, let us rewrite $g(\alpha, \beta)$ as
$$g(\alpha, \beta)=\sum_{i=1}^n e^{1+\alpha_i} \sum_{j=1}^n A_{i,j} e^{\beta_j}+\sum_{i=1}^n \alpha_i + \sum_{j=1}^n \beta_j.$$
We derive analytically a formula for $h(\beta) = \min_{\alpha \in \R^n} g(\alpha, \beta)$. Note that the problem of minimizing over $\alpha$ is separable and hence we just need to solve a sequence of optimization problems of the form
$$\min_{\alpha_i \in \R} ~~e^{-1-\alpha_i} \sum_{j=1}^n A_{i,j} e^{-\beta_j} + \alpha_i.$$
By a simple calculation, the optimal value of the above is $\log \inparen{\sum_{j=1}^n A_{i,j} e^{-\beta_j}}$ and hence we arrive at
$$ h(\beta)=  \sum_{i=1}^n \log \inparen{\sum_{j=1}^n A_{i,j} e^{-\beta_j}} + \sum_{j=1}^n \beta_j.$$
Further observe that $h(\beta) = \log \inparen{p_A\inparen{e^{-\beta_1}, e^{-\beta_2}, \ldots, e^{-\beta_n}}}+ \sum_{j=1}^n \beta_j.$ Hence in particular
$$\min_{\beta \in \R^n} h(\beta) = \min_{x>0} \log(p_A(x)) - \sum_{j=1}^n \log x_j= \min_{x>0}\log \frac{p_A(x)}{\prod_{j=1}^n x_j}.$$
Finally we obtain
$$R_E(A) = \min_{\alpha, \beta\in \R^n} g(\alpha, \beta) = \min_{\beta \in \R^n} h(\beta) = \log R_C(A).$$
\end{proof}
\section{Proof of Claim~\ref{lemma:poly}}\label{sec:proof_claim}

\begin{proof}
For any $x\in \R^{n\times n}_{>0}$ and any positive doubly stochastic matrix $B$ we have
\begin{align*}
\prod_{i=1}^n \sum_{j=1}^n x_{ij}M_{ij}&= \prod_{i=1}^n \sum_{j=1}^n B_{i,j} x_{i,j}\frac{M_{i,j}}{B_{i,j}}\\
&\geq \prod_{i=1}^n \prod_{j=1}^n \inparen{x_{i,j}\frac{M_{i,j}}{B_{i,j}}}^{B_{i,j}}\\
& =\prod_{1\leq i,j\leq n} \inparen{\frac{x_{i,j} M_{i,j}}{B_{i,j}}}^{B_{i,j}}
\end{align*}
The only inequality in the derivation above follows from applying Jensen's inequality to the function $\log$. We obtain
$$\inf_{x\in \R^{n\times n}_{>0}} \frac{r(x)}{\prod_{1\leq i,j\leq n} x_{i,j}^{B_{i,j}}} \geq \inparen{\frac{ M_{i,j}}{B_{i,j}}}^{B_{i,j}}.$$
To deduce equality it remains to observe that the applied Jensen's inequality is  tight whenever $x_{i,j}\frac{M_{i,j}}{B_{i,j}}=\mathrm{const.}$ which can be easily achieved by plugging in an appropriate value of $x$.
\end{proof}

\end{document}